\documentclass[prl,reprint,amsmath,amssymb,preprintnumbers,superscriptaddress,nofootinbib]{revtex4-1}

\usepackage{epsfig,epstopdf}
\usepackage{subfigure}
\usepackage{color}

\def\beq{\begin{equation}}
\def\eeq{\end{equation}}
\def\bea{\begin{eqnarray}}
\def\eea{\end{eqnarray}}

\newcommand{\lsim}{
\mathrel{\hbox{\rlap{\hbox{\lower4pt\hbox{$\sim$}}}\hbox{$<$}}}}
\newcommand{\gsim}{
\mathrel{\hbox{\rlap{\hbox{\lower4pt\hbox{$\sim$}}}\hbox{$>$}}}}

\newcommand{\dis}[1]{\begin{equation}\begin{split}#1\end{split}\end{equation}}
\newcommand{\mpl}{m_{\rm Pl}}

\begin{document}

\title{Thermodynamic interpretation of the de Sitter swampland conjecture}

\author{Min-Seok Seo}
\email{minseokseo57@gmail.com}
\affiliation{Department of Physics Education, Korea National University of Education,
\\ 
Cheongju 28173, Korea}

\begin{abstract}
\noindent 

 We interpret the de Sitter swampland conjecture in the thermodynamic point of view.
 When the number of degrees of freedom is enhanced as  the modulus rolls down the potential, the bound on $\mpl \nabla V/V$ is equivalent to the condition for the positive temperature phase. 
 The boundary between the positive- and the negative temperature phases is preferred by the classical system since the entropy density is maximized.
 The distance conjecture imposes that (quasi-)de Sitter spacetime is entirely in the negative temperature phase and statistically disfavored.
 By investigating the concave potential, we also justify the bound on $\mpl^2 \nabla^2 V/V$.

\end{abstract}
\maketitle

\section{Introduction}

 The instability of (quasi-)de Sitter (dS) spacetime has been a long standing issue. 
  While both primordial inflationary paradigm and $\Lambda$CDM model explaining the current accelerating Universe are well consistent with observations \cite{Aghanim:2018eyx}, constructing model based on string theory requires the nontrivial setup \cite{Dine:1985he}, as suggested by KKLT \cite{Kachru:2003aw} or large volume scenario \cite{Balasubramanian:2005zx}. 
  Once the string theory solution exists, apart from its naturalness, the anthropic principle \cite{Weinberg:1987dv} or the string landscape \cite{Bousso:2000xa} might account for the real world described by the solution.
  On the other hand, it was recently proposed that in any parametrically controllable regime (meta-)stable de Sitter spacetime is not allowed by quantum gravity \cite{Obied:2018sgi}.
 Among many conjectured effective field theory (EFT) properties consistent with quantum gravity (for  reviews, see, \cite{Brennan:2017rbf, Palti:2019pca}), this de Sitter swampland conjecture became controversial.
 Counter-examples have been studied through the Higgs and the axion potentials, from which the role of the curvature, not just the slope of the potential is emphasized \cite{Andriot:2018wzk,
  Denef:2018etk, Conlon:2018eyr, Murayama:2018lie, Choi:2018rze, Hamaguchi:2018vtv, Hebecker:2018vxz, Andriot:2018mav}.
  Such a situation calls for the refinement of the conjecture and also the physically acceptable arguments supporting it.
 In \cite{Garg:2018reu, Ooguri:2018wrx}, the refined dS swampland bound was formulated as
  \dis{&\mpl\frac{|\nabla V|}{V} \geq c,\quad\quad {\rm or}
  \\
  &\mpl^2\frac{{\rm min}(\nabla_i \nabla_j V)}{V} \leq -c'\label{Eq:dScond}}
  for some positive order one constants $c$ and $c'$ and more importantly, Bousso's covariant entropy bound \cite{Bousso:1999xy}, as well as the distance conjecture \cite{Ooguri:2006in} were considered to support the conjecture in \cite{Ooguri:2018wrx}.
   \footnote{The dS swampland bound can be written as the condition on the Hubble parameter $H$ instead of the potential  as given by \eqref{Eq:dScond} \cite{Seo:2018abc}. 
 This enables us to apply the dS swampland conjecture to exotic inflationary cosmology models, in which quasi-dS spacetime is not necessarily a consequence of the almost flat potential (see also \cite{Mizuno:2019pcm, Heckman:2018mxl, Heckman:2019dsj}).}
   That is, in the presence of some modulus  along which the number of physical degrees of freedom $N$ (hence the entropy) increases rapidly, the condition that the entropy cannot exceed the Gibbons-Hawking bound $S_{\rm GH}= \mpl^2 /H^2$ (we ignored the numerical factor of order one) results in the first bound in \eqref{Eq:dScond} as the Hubble parameter $H$ is also controlled by the modulus. 
 Meanwhile, the second bound states the breakdown of the semi-classical picture for dS spacetime by the tachyonic zero point quantum fluctuation.
   
 In this argument, the inequality itself comes from the thermodynamic property of spacetime, i.e., the entropy bound.
 On the other hand, the existence of the modulus with  properties given above as well as order one values of $c$ and $c'$ are results of the distance conjecture which claims that as the modulus traverses along the trans-Planckian geodesic distance towers of light degrees of freedom rapidly descend from UV.
  Motivated by these facts, in this letter we make a more systematic interpretation of the dS swampland conjecture using the language of thermodynamics.
 Moreover we try to specify the role of the distance conjecture which is irrelevant to thermodynamics.
  For this purpose, we take the ansatz $S_H=N^p (H/\mpl)^q$ for the entropy inside the horizon as considered in \cite{Ooguri:2018wrx}.
  The nonzero exponent $p$ stands for the effect from the large number of degrees of freedom as predicted by the distance conjecture while vanishing $p$ means that the entropy is purely geometric.
  
   From the thermodynamic definition of the temperature, we find that when the distance conjecture is applied to the modulus rolling down the potential, the bound on $\mpl |\nabla V|/V$  in \eqref{Eq:dScond} is in fact the condition for the  positive temperature phase, which is a consequence of the upper bound on the entropy under the EFT validity condition $H<\mpl$.
 Indeed, the boundary between the positive- and the negative temperature phases at which the bound on $\mpl |\nabla V|/V$ saturates has the maximal entropy density hence corresponds to a spacetime configuration preferred by the system.
 On the contrary, quasi-dS spacetime that belongs to the  negative temperature phase is not statistically favored.
 Finally, by estimating $\mpl^2\nabla^2V/V$ for the concave potential, we justify the second bound  in \eqref{Eq:dScond}.

\section{ The positivity condition on the temperature } 

To begin with, we recall that for (quasi-)dS spacetime, the entropy of the causally connected region cannot be arbitrarily large. 
 Let $S_H$ be the entropy within  the horizon of  size $H^{-1}$.
 Since the entropy of the system is maximized by that of the blackhole  of the same size which is given by the area law,  $S_H$ is bounded by $S_{\rm GH}=\mpl^2/H^2$ \cite{Gibbons:1977mu} (for a review of the holographic entropy bound including the area law, see, e.g., \cite{Bousso:2002ju}). 
 For the region larger than the horizon scale, the total entropy  is given by $S_H$ times the number of causally connected regions.
 Given the total volume  $a^3 v$, where $v$ is the constant comoving volume, there are  $n_H \equiv (a H)^3 v$ causally connected regions so the total entropy is given by $S=S_H n_H$ which cannot be larger than $S_{\rm GH} n_H$ \cite{Veneziano:1999ts}.

 Now consider  the ansatz for the entropy $S_H=N^p (H/\mpl)^q$  \cite{Ooguri:2018wrx}.
  Dividing the total entropy $S=(a^3v) N^pH^{q+3}/\mpl^q$ by the total volume ${\rm Vol}=a^3v$, we obtain the entropy density  $s=N^pH^{q+3}/\mpl^q$.
We require  $p\ge 0$ since negative $p$ is not consistent with our intuition that the larger the number of degrees of freedom, the larger the entropy.
From the first law of thermodynamics, $dE=-pd({\rm Vol})+TdS$, the temperature is defined as $T^{-1}=(\partial S/\partial E)_{\rm Vol}=(\partial s/\partial \rho)_{\rm Vol}$.
 Using the fact that $H$ depends on $\rho$ through
 \dis{3\mpl^2 H^2 = \rho \simeq V(\phi),\label{Eq:eom}} 
  and $N$ depends on $\rho\simeq V(\phi)$ through $\phi$, we obtain
 \dis{&\frac{\partial H^{q+3}}{\partial \rho}\Big|_{\rm Vol}=\frac{1}{3\mpl^2}\frac{d (H^2)^{\frac{q+3}{2}}}{d H^2}=\frac{q+3}{6\mpl^2}H^{q+1},
 \\
 &\frac{\partial N^p}{\partial \rho}\Big|_{\rm Vol}=\frac{d \phi}{d V}\frac{d N^p}{d \phi}=p\frac{N^p}{V'}\frac{1}{N}\frac{d N}{d\phi}
,}
 from which $T^{-1}$ is given by
 \dis{T^{-1}=\Big[\frac{q+3}{6}+{\rm sgn}(V')\frac{p}{3\sqrt{2 \epsilon_V}}\frac{\mpl}{N}\frac{d N}{d\phi}\Big]N^p\frac{H^{q+1}}{\mpl^{q+2}}.\label{Eq:T}}
 Here, $\mpl(V'/V)$ is denoted as ${\rm sgn}(V')\sqrt{2\epsilon_V}$, following the conventional definition of the slow-roll parameter in  inflationary cosmology.
 
 We note here that since $H<\mpl$ for the EFT we are working with to be valid, for $N> 1$ and $p> 0$, the entropy bound $S_H \leq S_{\rm GH}$  or $N^p\leq(H/\mpl)^{-2-q}$ holds only if $q>-2$.
 When $p=0$, since the EFT validity condition and the entropy bound just give $(H/\mpl)^{-2-q} \geq 1$, we have $q \geq-2$ \cite{Brustein:1999ua}. 
 In any of two cases, the first term in \eqref{Eq:T} is always positive.
  
  On the other hand, for $p=0$  the second term vanishes hence the temperature is always positive. 
  When $p\ne 0$, the sign of the nonzero second term in \eqref{Eq:T} depends on  the shape of the potential.
 To see this, we first choose the fiducial value of the modulus $\phi_0$.
Then the distance conjecture predicts that  the rate $(\mpl/N)dN/d\phi$ at $\phi$ is positive as $\phi$ recedes from $\phi_0$, with $|(\mpl/N)dN/d\phi|\sim {\cal O}(1)$.
 Therefore, ${\rm sgn}(V')({\mpl}/{N})({dN}/{d\phi})$ becomes {\it negative} ({\it positive}) if $\phi$ gets away from $\phi_0$ by {\it rolling down} ({\it climbing up}) the potential.

  As a specific example, we can consider the potential which has a local maximum or a local minimum.
 When $V(\phi_0)$ is  the local maximum (minimum) of the potential, ${\rm sgn}(V')({\mpl}/{N})({dN}/{d\phi})$ is always negative (positive). 
 Especially, the string theory models for the meta-stable dS spacetime as a local minimum like KKLT \cite{Kachru:2003aw} or large volume scenario \cite{Balasubramanian:2005zx} realize the positive definite ${\rm sgn}(V')({\mpl}/{N})({dN}/{d\phi})$.
  We note that whether the potential changes the direction or maintains the flatness at the local maximum of the potential  depends on the value of $\mpl^2\nabla^2 V/V$. 
 The bound on $\mpl^2\nabla^2 V/V$ in \eqref{Eq:dScond} asserts that the latter is excluded by quantum gravity.

 In terms of inflationary cosmology, the Friedmann equations \eqref{Eq:eom} and $2\mpl^2\dot{H}=-(\rho+p)$, together with the relations $\rho=\frac12\dot{\phi}^2+V$ and $p=\frac12\dot{\phi}^2-V$ impose   $\dot{H}<0$ such that the slow-roll parameter satisfies the relation $\epsilon_H=-\dot{H}/H^2=\dot{\phi}^2/(2\mpl^2H^2)$.
 If we take $\dot{\phi}>0$ as $\phi$  gets farther from the initial value $\phi_0$, the time derivative of \eqref{Eq:eom} results in sgn$(V')=-1$.
 The distance conjecture ensures that $(\mpl/N)dN/d\phi$ is positive so we have ${\rm sgn}(V')({\mpl}/{N})({dN}/{d\phi})<0$.
 On the other hand, having ${\rm sgn}(V')({\mpl}/{N})({dN}/{d\phi})>0$  through ${\rm sgn}(V')=+1$ and $({\mpl}/{N})({dN}/{d\phi})>0$ at the same time is nontrivial  since $-\dot{H}$ (hence $\rho+p$) is no longer positive.
  It violates the null energy condition which the bound for $\mpl \nabla V/V$ in \eqref{Eq:dScond} is regarded as being based on \cite{Obied:2018sgi}.
  Moreover, the energy can be unstable by, e.g., negative kinetic term as can be found in the phantom model \cite{Caldwell:1999ew} (see also \cite{Carroll:2003st}).

 When ${\rm sgn}(V')({\mpl}/{N})({dN}/{d\phi})>0$, the temperature is always positive definite but lower than $(6/(q+3))N^{-p}{\mpl^{q+2}}/{H^{q+1}}$.
 On the contrary, in the case of ${\rm sgn}(V')({\mpl}/{N})({dN}/{d\phi})<0$, too large value of the second term in \eqref{Eq:T} drives  the temperature negative.
 Then the positivity condition on the temperature  sets the bound on $\epsilon_V$ as
 \dis{\epsilon_V^{1/2} \geq \frac{\sqrt2 p}{q+3}\Big|\frac{\mpl}{N}\frac{dN}{d\phi}\Big| \equiv \epsilon_c^{1/2},\label{Eq:dSbound1}} 
  which is what \cite{Ooguri:2018wrx} obtained as the first bound in \eqref{Eq:dScond}.

 \begin{figure}[!ht]
  \begin{center}
   \includegraphics[width=0.35\textwidth]{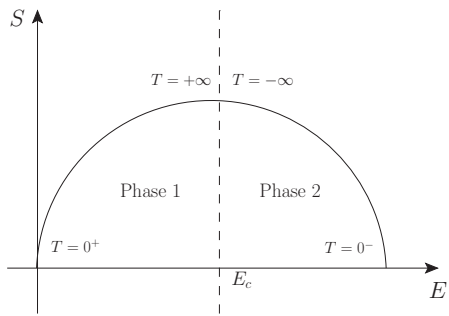}
  \end{center}
 \caption{The entropy-energy relation for the system having the positive- and the negative temperature phases.
 The temperature diverges when the entropy is bounded from above.
  }
\label{fig:NegT}
\end{figure}

\section{ The negative temperature phase } 
 
 So far we find that the bound on $\mpl(V'/V)$ as given by \eqref{Eq:dSbound1} is interpreted as the positivity condition on the temperature provided ${\rm sgn}(V')({\mpl}/{N})({dN}/{d\phi})<0$.
 On the other hand, a negative temperature is not an unphysical quantity, whereas it does not appear in the usual classical thermodynamic system.
A negative temperature just means that the entropy decreases as the energy increases at fixed volume, which appears when the possible energy of the system has an upper bound.
 In this case, the number of microstates occupied by the system does not diverge at high energy.
 In Fig. \ref{fig:NegT}, we depict the entropy-energy relation for the system  having the positive (phase 1) and the negative (phase 2) temperature phases. 
 We infer that the system has a zero temperature if the whole system is either on the ground state ($T=0^+$) or on the highest energy state ($T=0^-$).
 We also expect that when the entropy has a maximum at $E=E_c$, the temperature becomes infinity at $E_c$ by $T^{-1}(E_c)=\partial S/\partial E|_{E_c}=0$.
 Then  $E_c$  characterizes the boundary between the positive- and the negative temperature phases  giving $T(E_c^-)=+\infty$ and $T(E_c^+)=-\infty$. 
 The system at a negative temperature has rather higher energy compared to that at a positive temperature.

 \begin{figure}[!ht]
  \begin{center}
   \includegraphics[width=0.40\textwidth]{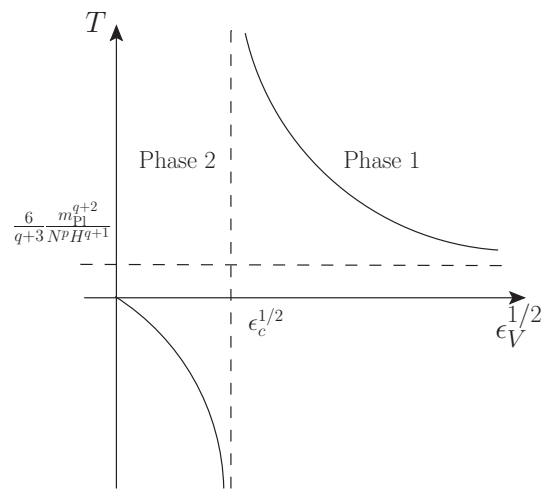}
  \end{center}
 \caption{The phase diagram for quasi-dS spacetime satisfying ${\rm sgn}(V')({\mpl}/{N})({dN}/{d\phi})<0$ for fixed $N$ and $H$.
 The temperature diverges at $\epsilon_V=\epsilon_c$, in which the entropy is maximized.
  }
\label{fig:Phase}
\end{figure}
 
 It is remarkable that for ${\rm sgn}(V')({\mpl}/{N})({dN}/{d\phi})<0$, the behavior of the temperature of (quasi-)dS spacetime given by  \eqref{Eq:T}  follows that of the system described above, as shown in Fig. \ref{fig:Phase}. 
 We note that whereas we have considered the entropy dependence on the energy in the previous discussion, Fig. \ref{fig:Phase} describes the  entropy  variation with respect to $\epsilon_V$ while  $H$ (hence the energy density) as well as $N$ are fixed.
 Indeed, considering the change in $V$ under the infinitesimal variation $\Delta \phi$ along the direction of  $(\mpl/N)dN/d\phi>0$, as $\epsilon_V$ gets larger we have the steeper potential so the height of the potential  soon decreases.
  Then qualitatively  the average energy density over $\Delta \phi$ gets smaller for larger $\epsilon_V$.
  It is maximized for $\epsilon_V=0$ in which $H$ is maintained at a constant value.

We can also understand the  entropy dependence on $\epsilon_V$ in the following way.
Since the entropy density is given by $s=N^p H^{q+3}/\mpl^q$ with both $p$ and $q+3$ positive,  for the potential satisfying  ${\rm sgn}(V')({\mpl}/{N})({dN}/{d\phi})<0$ we have a tension between growing $N$ and diminishing $V$ along $\phi$, as each of them raises and lowers the entropy density respectively.
  If the potential is nearly flat, i.e., $\epsilon_V\simeq 0$ the decrement of the potential is very small whereas  $|(\mpl/N)(dN/d\phi)|$ is order one so the entropy density tends to increase along $\phi$ by the growing $N$ effect.
  But such a growth of the entropy density is limited as the entropy  is bounded by $S_{\rm GH}$. 
 Therefore,  for large $\epsilon_V$ the decrease rate of $V$ becomes dominant and the entropy density diminishes along $\phi$.

 The argument above shows that $\epsilon_V$ determines whether the entropy density is on the way of increase or decrease along $\phi$ (which is in the diminishing $V\simeq \rho$ direction) at given $H$ (or the height of the potential).  
 Then we expect that there exists a value of $\epsilon_V$ which in fact is given by $\epsilon_c$ (in \eqref{Eq:dSbound1}) that the effect of the increasing $N$ becomes strongest.
 When the potential with sgn$(V')=-1$ has a slope giving $\epsilon_c$ at some specific value of $H$, the entropy density is maximized there.
 This point corresponds to the boundary between the positive- and the negative phases so we have $T(\epsilon_c^-)=-\infty$ and $T(\epsilon_c^+)=+\infty$. 
  We note that the divergence of the temperature at the boundary is a result of the cancellation between the positive first term and the negative second term in \eqref{Eq:T}, and the positivity of the first term is ensured by the Gibbons-Hawking entropy bound together with the EFT validity condition $H<\mpl$.
 Such a configuration with the maximum entropy is statistically most probable as the number of microstates giving it is the largest there. 
 If $(\mpl/N)dN/d\phi\ll {\cal O}(1)$, spacetime at the boundary is close to dS so we can find quasi-dS spacetime with $\epsilon_V \ll 1$  preferred by the system. 
 However, the distance conjecture imposes  $(\mpl/N)dN/d\phi\sim {\cal O}(1)$, so  spacetime at the boundary strongly deviates from perfect dS. 
 Then entire quasi-dS spacetime is in the negative temperature phase and no longer preferred by the system : it has too small entropy.
 We also note that   as $\epsilon_V\to \infty$ the temperature approaches to  the asymptotic value $(6/q+3)N^{-p}{\mpl^{q+2}}/{H^{q+1}}$ instead of zero implying the absence of the ground state. 
  But this is not physical since spacetime severely deviates from quasi-dS, so \eqref{Eq:eom} is no longer a good description : kinetic energy we have ignored becomes important.
 
 Such a behavior of the entropy density with respect to $\epsilon_V$ imposes the bound on $\mpl^2\nabla^2V/V$ when applied to the concave potential.
 For the concave potential having a maximum at $\phi=\phi_0$, as we move away from the top of the potential $\epsilon_V$ gradually increases from $0$.
 Suppose at some value of $\phi$, which we call $\phi_c$, $\epsilon_V$ reaches  $\epsilon_c$, which implies that the entropy density is maximized at $H(\phi_c)$.
Since the increase of the entropy is a result of  the rapidly growing $N$ which is  driven by the trans-Planckian excursion of the modulus as expected from the distance conjecture, we estimate $\phi_c-\phi_0\sim \mpl$. 
 From this the change in $V'$ over the range $[\phi_0, \phi_c]$ is gives
  \dis{\frac{\mpl^2}{V(\phi_0)}\frac{\Delta V'}{\Delta \phi}& \simeq \frac{\mpl^2}{V(\phi_0)}\frac{V'(\phi_c)-V'(\phi_0)}{\phi_c-\phi_0}
  \\
  &\simeq {\rm sgn}(V')\Big[\frac{V(\phi_c)}{V(\phi_0)}\sqrt{2\epsilon_c}-\sqrt{2\epsilon_V}(\phi_0)\Big]
  \\
  &= -\frac{2p}{q+3}\frac{V(\phi_c)}{V(\phi_0)}\Big|\frac{\mpl}{N}\frac{dN}{d\phi}\Big|.}
  Since $\epsilon_V$ is not large enough to dominate over the growing $N$ effect, it is reasonable to assume that $V(\phi_c)/V(\phi_0)$ is not so much smaller than $1$.
  As the absolute value of $\mpl^2V''/V$ is maximized at the top of the concave potential, we conclude that near the top of the potential at which $\epsilon_V \ll 1$,
  \dis{\mpl^2\frac{V''}{V} \lesssim -\frac{2p}{q+3}\frac{V(\phi_c)}{V(\phi_0)}\Big|\frac{\mpl}{N}\frac{dN}{d\phi}\Big|,}
  as the second bound in \eqref{Eq:dScond} states.

 \section{ Conclusion }

 So far, we present the thermodynamic interpretation of the dS swampland conjecture for the potential satisfying ${\rm sgn}(V')(\mpl/N)(dN/d\phi)<0$ under the distance conjecture.
 Such a potential corresponds to the system having the negative temperature phase.
 When the bound on $\mpl\nabla V/V$ saturates the system has a maximum entropy and the value of $\epsilon_V$ at this point is constrained by  the Gibbons-Hawking entropy bound, the EFT validity condition as well as the distance conjecture.
 On the other hand, our thermodynamic consideration does not tell us about the (quasi-)dS vacuum as a local minimum around which the potential does not have any direction satisfying ${\rm sgn}(V')(\mpl/N)(dN/d\phi)<0$.
 As this case includes the string theory model for the metastable dS vacuum, it would be meaningful to find out the thermodynamic argument supporting or avoiding the dS swampland bound on it.

\vspace{5mm}
\begin{acknowledgments}

Acknowledgments: MS thanks Jinn-Ouk Gong for comments.

\end{acknowledgments}


\end{document}